\documentclass[aps,prb,reprint,amsmath,amssymb,showpacs,nobibnotes]{revtex4-1}

\usepackage{graphicx}
\usepackage{bm}     %
\usepackage{hyperref} %
\usepackage{xcolor}
\usepackage{braket}

\hypersetup{
    colorlinks,
    linkcolor={blue!80!black},
    citecolor={blue!80!black},
    urlcolor={blue!80!black}
}
\usepackage{dcolumn} %
\usepackage{dsfont}    %
\usepackage{color}
\usepackage{textcomp,gensymb} %

\makeatletter
 \begingroup
  \catcode`\_=\active
  \protected\gdef_{\@ifnextchar|\subtextup\sb}
 \endgroup
\def\subtextup|#1|{\sb{\textup{#1}}}
\AtBeginDocument{\catcode`\_=12 \mathcode`\_=32768}
\makeatother

\begin{document}

\title{NOON-state interference in the frequency domain}

\author{Dongjin Lee}
\affiliation{Department of Physics, Pohang University of Science and Technology (POSTECH), Pohang, 37673, South Korea}

\author{Woncheol Shin}
\affiliation{Department of Physics, Pohang University of Science and Technology (POSTECH), Pohang, 37673, South Korea}

\author{Sebae Park}
\affiliation{Department of Physics, Pohang University of Science and Technology (POSTECH), Pohang, 37673, South Korea}

\author{Junyeop Kim}
\affiliation{Department of Physics, Pohang University of Science and Technology (POSTECH), Pohang, 37673, South Korea}

\author{Heedeuk Shin}
\email{heedeukshin@postech.ac.kr}
\affiliation{Department of Physics, Pohang University of Science and Technology (POSTECH), Pohang, 37673, South Korea}

\date{\today}

\begin{abstract}
The examination of entanglement across various degrees of freedom has been pivotal in augmenting our understanding of fundamental physics, extending to high dimensional quantum states, and promising the scalability of quantum technologies. In this paper, we demonstrate the photon number path entanglement in the frequency domain by implementing a frequency beam splitter that converts the single-photon frequency to another with 50\% probability using Bragg scattering four-wave mixing. The two-photon NOON state in a single-mode fiber is generated in the frequency domain, manifesting the two-photon interference with two-fold enhanced resolution compared to that of single-photon interference, showing the outstanding stability of the interferometer. This successful translation of quantum states in the frequency domain will pave the way toward the discovery of fascinating quantum phenomena and scalable quantum information processing. 
\end{abstract}

\maketitle

\section*{Introduction}
Photonic entanglement plays a crucial role in resolving fundamental questions in quantum mechanics and exploiting quantum information technology's merits~\cite{flamini2018photonic}. Extensive investigations have been conducted into entanglement in various degrees of freedom, including polarization~\cite{hamel2014direct}, path~\cite{silverstone2015qubit}, orbital angular momentum~\cite{mair2001entanglement}, and time-bin~\cite{kim2022quantum}. These studies have been key in investigating diverse quantum phenomena, enhancing the performance of quantum communications~\cite{nadlinger2022experimental, duan2001long}, protecting quantum states from environment noise~\cite{kim2021noise}, and surpassing classical limitations in metrology~\cite{nagata2007beating, shin2011quantum, shin2013enhancing, polino2020photonic}. 

The NOON state, denoted as $\ket{\psi} = (\ket{N}_1 \ket{0}_2+ \ket{0}_1 \ket{N}_2) / \sqrt{2}$, is commonly referred to as the photon number path entangled state. This state represents a superposition between $N$ photons in path 1 and zero photons in path 2 and vis-versa. A distinctive feature of the NOON state is its N-fold enhancement in phase sensitivity, which allows it to surpass the limitations of classical light measurement~\cite{kok2002creation, nagata2007beating}. This attribute has made the NOON state a fundamental resource in quantum applications, including quantum lithography~\cite{shin2011quantum}, quantum imaging~\cite{israel2014supersensitive}, and quantum metrology~\cite{shin2013enhancing, hong2021quantum}. Moreover, unique entanglement properties of the NOON state pave the way toward the exploration of diverse topics in quantum information science, such as nonlocality~\cite{teh2016signifying}, quantum error correction~\cite{bergmann2016quantum}, and tight-binding model~\cite{lebugle2015experimental, bromberg2010bloch}.

Recently, the focus on quantum states in the frequency domain has intensified due to its potential for high-dimensional state extensibility, spatial single-mode propagation, stability, miniaturization, and compatibility with fiber networks~\cite{joshi2018frequency, clemmen2016ramsey, lu2022bayesian, lu2018electro, clemmen2018all}. The potential benefits have spurred extensive research into a variety of techniques for photon creation and manipulation within the frequency domain~\cite{joshi2022picosecond, joshi2018frequency, lu2022bayesian, lu2018electro, kues2017chip, clemmen2018all}. Furthermore, frequency-domain classical light controls have facilitated the demonstration of complex physical phenomena, including three-dimensional photonic topological insulator~\cite{lin2018three} and complex long-range coupling~\cite{bell2017spectral, wang2020multidimensional}.  

In this work, we demonstrate the NOON-state interference in the frequency domain for the first time, which is a crucial resource in quantum optics, to the best of our knowledge. A quantum frequency translation process acts like a 50:50 beam splitter with about a 50\% probability of converting a single-photon frequency to another and enables the creation of the $N=2$ NOON state in the frequency domain using a nondegenerate photon pair. The state is subsequently reintroduced into the frequency beam splitter through reflection, and the relative phase of the NOON state is controlled via a variable delay line. We observe two-photon bunching and anti-bunching effects in the frequency domain against the relative phase, with the oscillation frequency of the NOON state displaying a two-fold enhancement compared to single-photon interference. Furthermore, our approach ensures an extremely stable interferometer due to the single-mode propagation of two-color components, even without any stabilization method. Consequently, our work represents a significant step toward exploring novel quantum effects and facilitating new tools for quantum information processing.

\section*{Results}
\begin{figure}[t!]
\centering
\includegraphics[width=3.35in]{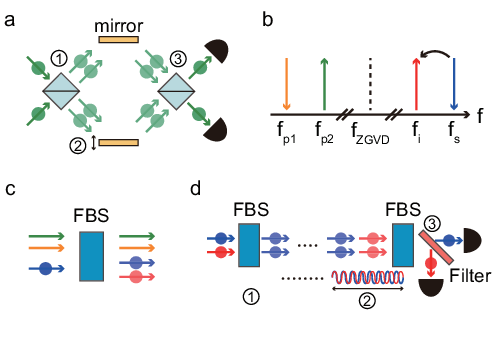}
\caption{\textbf{Schematic diagram.} \textbf{a}, Traditional NOON-state interference (N = 2) scheme in the spatial domain. \textbf{b}, BS-FWM frequency diagram. $f_|xx|$ ($\rm xx= p1, p2, s, i, ZGVD$) indicates the frequencies of two pump fields, signal photon, idler photon, and the zero-group-velocity dispersion (ZGVD) of a nonlinear medium, respectively. \textbf{c}, Schematic diagram of a frequency beam splitter through the BS-FWM process with two pump fields, an input photon, and a target photon. FBS: frequency beam splitter. \textbf{d}, Proposed NOON-state interference scheme in the frequency domain.}
\label{fig1}
\end{figure}

Fig.~\ref{fig1} illustrates our experimental diagram. The traditional NOON-state interference in the spatial domain is shown in Fig.~\ref{fig1}\textbf{a}. Here, \textcircled{1} two indistinguishable single photons are simultaneously introduced into two input modes of a beam splitter, yielding the Hong-Ou-Mandel (HOM) effect and creation of a NOON state ($\rm N=2$). \textcircled{2} We can control the relative phase between the two paths by sliding a mirror, and \textcircled{3} the NOON state can be combined by the second beam splitter. The resultant state exhibits two-photon bunching and anti-bunching effects in the path modes, contingent on the relative phase. The bunching cycles of the NOON state (N = 2) display a two-fold enhancement compared to single-photon interference.

In this study, we generate the NOON state in the frequency domain using optical elements that serve functions analogous to those in traditional methods. We utilize frequency beam splitters based on quantum frequency translation, akin to the spatial beam splitters. The frequency translation method used in this study is BS-FWM~\cite{lee2023translation, joshi2018frequency, clemmen2016ramsey}, which facilitates the simultaneous annihilation of an input photon and the creation of a target photon, driven by two classical pump fields~\cite{mckinstrie2005translation, mcguinness2010quantum}, as seen in Fig.~\ref{fig1}\textbf{b}. From now on, we denote the input and target modes as signal and idler modes, respectively. The idler photon's frequency obeys the energy conservation and phase-matching conditions, which can be fulfilled by symmetrically placing the four fields around the ZGVD frequency of the nonlinear material. The idler photon's frequency is determined by $f_{\rm{i}} = f_{\rm{s}} + f_{\rm{p1}} - f_{\rm{p2}}$, where $f_{\rm{l}}$ ($\rm l = i, s, p1, p2$) represents the frequency of the idler, signal, pump1, and pump2, respectively. Vertical arrows pointing upwards signify the creation process, while those pointing downwards represent the annihilation process. The evolution of the annihilation mode operators via BS-FWM is dictated by~\cite{mckinstrie2005translation, mcguinness2010quantum, clemmen2016ramsey}

\begin{equation}
\label{eq1}
    \begin{aligned}
        \hat{a}_{\rm{s,out}} &= \cos(gL) \hat{a}_{\rm{s,in}} + ie^{i\phi} \sin(gL) \hat{a}_{\rm{i,in}}
    \end{aligned}
\end{equation}
\begin{equation}
\label{eq2}
    \begin{aligned}
        \hat{a}_{\rm{i,out}} &= ie^{-i\phi} \sin(gL) \hat{a}_{\rm{s,in}} + \cos(gL) \hat{a}_{\rm{i,in}}
    \end{aligned}
\end{equation}

\noindent
where $\hat{a}_{\rm{m,n}}$ ($\rm m = s, i$, $\rm n = in, out$) denotes the annihilation operators of signal and idler modes in the input and output modes, respectively. $L$ symbolizes the length of the nonlinear medium, and $\phi$ signifies the phase difference between the two pump beams. The parameter $g$ is defined as $g \equiv \gamma P$, where $\gamma$ is the nonlinear coefficient and $P_1$ and $P_2$ are the powers of the pump beam 1 and 2, respectively. In this study, $P$ is the total power of the two pump beams, which are set to be equal ($P/2 = P_1 = P_2$). This ensures the attainment of the phase-matching condition, independent of the total pump power~\cite{mckinstrie2005translation}. Eq.~(\ref{eq1}) and Eq. (\ref{eq2}) equate to operators describing the functionality of a frequency beam splitter. A frequency beam splitter allows frequency translation of a signal photon to the desired idler frequency with a translation probability, as shown in Fig.~\ref{fig1}\textbf{c}. For an initial insertion of a signal portion, the probability of persisting in the signal frequency, similar to transmittance in a spatial beam splitter, is $\cos^2(gL)$, while that of translating to the idler frequency, like reflectance in a spatial beam splitter, is $\sin^2(gL)$. Control of the splitting ratio in the frequency domain is achievable by changing the BS-FWM pump power. Consequently, the BS-FWM effect enables the implementation of frequency beam splitters analogous to spatial beam splitters~\cite{clemmen2016ramsey, kobayashi2016frequency}. In this work, we have taken the frequency beam splitter a step further by creating the NOON state in the frequency domain, which is a significant milestone in quantum optics, for the first time to the best of our knowledge.

Our proposed concept of NOON-state interference in the frequency domain is illustrated in Fig.~\ref{fig1}\textbf{d}. \textcircled{1} Instead of feeding two identical single photons into the two input ports of a spatial beam splitter, we introduce two single photons — one embodying the signal frequency, the other the idler frequency — into a frequency beam splitter. Here, different frequencies entering the frequency beam splitter correspond to the two input ports of the spatial beam splitter. A translation probability of 50\% induces the HOM effect in the frequency domain, and the resultant NOON state of N = 2 in the frequency domain signifies the frequency two-photon bunching either in the signal frequency or the idler frequency. \textcircled{2} We control the relative phase of the NOON state by adjusting the position of the secondary frequency beam splitter as the length difference between the signal and idler wavelengths causes a relative phase shift between them. \textcircled{3} The final state, after traversing the secondary frequency beam splitter, exhibits periodic two-photon bunching and anti-bunching effects in the frequency domain, influenced by the relative phase. The bunching effect can be observed by placing a filter to separate the signal and idler photons. Therefore, we are able to implement the NOON-state interference in the frequency domain via Bragg scattering four-wave mixing.

In the interferometer depicted in Fig.~\ref{fig1}\textbf{d}, the output state after the second frequency beam splitter is given by

\begin{equation}
\label{eq3}
\begin{aligned}
\ket{\psi_|NOON|} &= \frac{1}{\sqrt{2}} e^{i(\Delta \phi + \phi)} \sin(\Delta \phi) \ket{2}_|s| \ket{0}_|i| \\
&- \frac{1}{\sqrt{2}} e^{i(\Delta \phi - \phi)} \sin(\Delta \phi) \ket{0}_|s| \ket{2}_|i| \\
&- e^{i\Delta \phi} \cos(\Delta \phi) \ket{1}_|s| \ket{1}_|i|,
\end{aligned}
\end{equation}

\noindent
where $\ket{N}_|m,out|$ ($\rm m = s, i$) represent the Fock states ($N$-photon-number states) at the output signal and idler modes, respectively. The relative phase $\Delta \phi$ is defined as $\Delta \phi = 2\pi \Delta f L/c$, where $\Delta f$ is the frequency difference between the signal and idler modes ($\Delta f = f_|s| - f_|i|$). The parameter $L$ represents the distance between the two frequency beam splitters. This formulation reveals the oscillating two-photon bunching probability with a period of $c/(2\Delta f)$. In contrast, when classical light or single photons are injected into the same interferometer, the oscillation period doubles, becoming $c/\Delta f$. Notably, while the coefficient $\phi$ in Eqs.~(\ref{eq1}) and (\ref{eq2}) is necessary for the beam splitter operators, it does not influence the measurement outcomes for the HOM effect, NOON-state interference, or single-photon interference in the frequency domain. See Supplementary Information, Sec.~I, for details about analytic calculations.

\begin{figure*}[t!]
\centering
\includegraphics[width=6.7in]{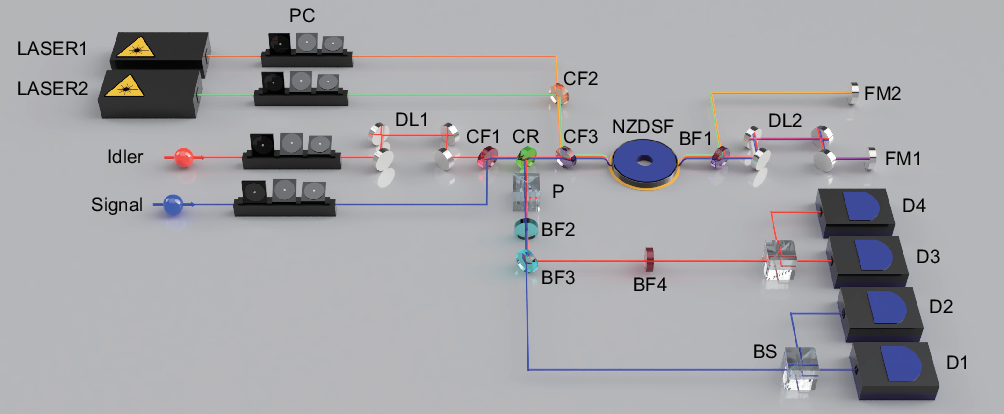}
\caption{\textbf{Experimental setup of the NOON-state interference.} PC: polarization controller, DL: delay line, CF: combining filter, CR: circulator, NZDSF: non-zero dispersion-shifted fiber, BF: bandpass filter, FM: Faraday mirror, P: polarizer, BS: beam splitter, D: superconducting nanowire single-photon detector.
}
\label{fig2}
\end{figure*}

\begin{figure}[t!]
\centering
\includegraphics[width=3.35in]{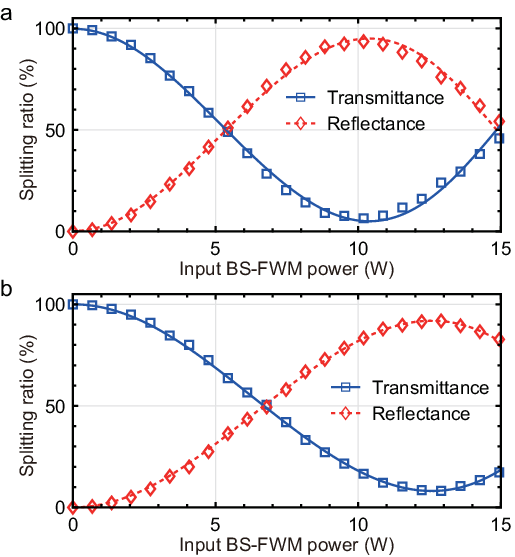}
\caption{\textbf{Splitting ratio in a frequency beam splitter for forward and backward propagation.} 
The variation in splitting ratios as a function of input BS-FWM pump power is depicted for \textbf{a} forward and \textbf{b} backward propagation. The experimentally measured transmittance and reflectance are presented by blue squares and red diamonds, respectively. The fitting of these data points is achieved through solid blue lines for transmittance and dashed red lines for reflectance, both of which are modeled as envelopes of sinusoidal functions.
}
\label{fig3}
\end{figure}

Our experimental setup for observing NOON-state interference in the frequency domain is shown in Fig.~\ref{fig2}. Non-degenerate photon pairs are created via spontaneous four-wave mixing in a 200-m-long single-mode fiber (Corning, SMF-28). The pump laser has a center wavelength of 1269.50 nm and temporal duration of 0.1 ns. Details about the photon-pair generation is described in the Methods section. The signal and idler photons have center wavelengths of 1267.89 nm and 1271.11 nm, respectively, with a bandwidth of about 0.7 nm. After combining them with a combining filter (CF1), they are introduced to the frequency beam splitter. The relative delay between them is controlled using a delay line (DL1).

The two BS-FWM pump lasers (LASER1 and LASER2) have wavelengths of 1551.16 nm (193.27 THz) and 1546.36 nm (193.87 THz), yielding a translation frequency of 600 GHz, and both of them have a repetition rate of 20 MHz and a temporal duration of 0.5 ns. After combining the two BS-FWM pump lasers using a combining filter (CF2) and equalizing their power, they are synchronized with the photon-pair generation pump electronically. The generated photon pairs and BS-FWM pumps are directed through a 100-m-long non-zero dispersion-shifted fiber (NZDSF) with a ZGVD wavelength of 1401 nm, where the BS-FWM effect with 50\% conversion efficiency serves as a frequency beam splitter, facilitating the HOM effect in the frequency domain. The temporal overlaps and polarization direction between the pair photons and BS-FWM pumps are optimized. After the first pass through the frequency beam splitter, the two photons possess identical frequencies, either at the signal or idler frequency, creating the $N=2$ NOON state in the frequency domain. The photon pairs and BS-FWM pumps are separated by a bandpass filter (BF1) and reflected back by Faraday mirrors (FM1 and FM2), enabling their backward propagation through the NZDSF. Additionally, we adjust the optical path length (DL2) to ensure temporal overlap between the pair photons and BS-FWM pumps during the backward propagation.

The relative phase within the NOON state is controlled by a delay line (DL2). Note that both signal and idler photons traverse a single-mode fiber. Given the wavelength discrepancy between the signal (1267.89 nm) and idler (1271.11 nm) photons, which is approximately 3.22 nm, a path length variation equivalent to the idler wavelength would generate a relative phase shift of approximately 0.253\% between the signal and idler. A 0.5-mm path length variation to DL2 corresponds to one idler wavelength of the relative phase shift. This path length variation is identical to a temporal change of 1.67 ps, much smaller than the pump duration of BS-FWM, guaranteeing still good temporal overlap between the BS-FWM pumps and the pair photons. Under our experimental conditions, where the pair photons pass the DL2 twice, a delay-line change of about 0.25 mm would cause one period in a classical interference pattern.

A circulator (CR) extracts the backward photons from the NOON-state interferometer. BS-FWM pumps are initially filtered out by a combining filter (CF3), and any residual pumps are further suppressed by a bandpass filter (BF2) with a bandwidth of 16.9 nm centered at 1270 nm, achieving a rejection rate exceeding -120 dB. The signal and idler photons are separated by additional bandpass filters (BF3 and BF4), each with a bandwidth of 0.7 nm centered at 1267.89 nm and 1271.11 nm, respectively. At each path, a beam splitter along with two superconducting nanowire single-photon detectors (SNSPDs; Scontel, HED model) facilitate the post-selection of two-photon events. Given that the SNSPDs are optimized for C-band photons, their measurement efficiency diminishes to around 40\% in the O-band. Furthermore, their dark count rates are maintained around 100 Hz. The data is collected via a time-correlated single photon counting (TCSPC, Swabian instruments) module with a coincidence window of 0.3 ns. The total transmission of the quantum frequency translation setup is about 59\% (-2.3 dB), including the transmissions from the combining filter (CF3: 93\%, -0.3 dB), NZDSF and bandpass filter (NZDSF and BF1: 87\%, -0.6 dB), and noise block filter for the BS-FWM effect (BF2: 72\%, -1.4 dB). Note that the intrinsic transmission from a 100-m-long NZDSF is negligible ($\sim$ 99\%, -0.05 dB).

In our study, we meticulously characterize the splitting ratios of a frequency beam splitter for both the forward and backward directions. The splitting ratio can be determined by monitoring the converted and non-converted signal single-photon counts while we isolate the idler photons. First, we ensure BS-FWM operates in a single direction, achieved by carefully adjusting the optical delay (DL2) and BS-FWM pump arrival time. The DL2 is varied by about 4.5 ns from its optimum position for both the forward and backward BS-FWM. To activate BS-FWM only in the forward direction, the pumps and signal photons arrive at the frequency beam splitter at their first passing. Due to the double passing of the 4.5-ns delay, the photons and BS-FWM pumps are separated by 9.0 ns, which is 18 times longer than the BS-FWM pump duration, causing no BS-FWM in backward propagation. Similarly, to achieve only backward BS-FWM, we postpone the BS-FWM pumps electronically by 9.0 ns. Then, the pumps coincide with signal photons at the frequency beam splitter at the second passing.

The measured splitting ratio in the forward and backward direction is shown in Fig.~\ref{fig3}. The transmittance (blue squares, $T$) of the splitting ratio is defined as the ratio between non-converted photon counts and total (converted and non-converted) photon counts at each pump power. As we assume that the frequency translation occurs only between signal and idler wavelengths, the reflectance (red diamonds) of the splitting ratio is $R = 1 - T$. In reality, a small percentage of photons are diverted due to scattering into higher-order modes, but this loss is minor and does not significantly affect the calculation of the splitting ratio~\cite{lee2023translation}.

\begin{figure}[t!]
\centering
\includegraphics[width=3.35in]{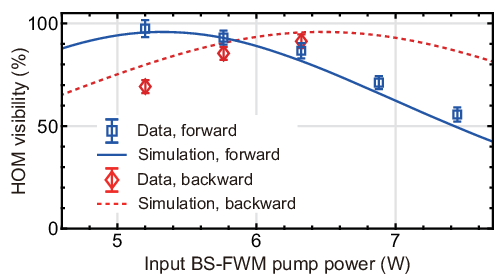}
\caption{\textbf{HOM interference using the frequency beam splitter.} Blue squares and red diamonds represent the net visibilities for the forward and backward directions of the HOM effect, respectively. The error bars are calculated from the fit coefficients and confidence bounds of the HOM interference. Solid blue and red dashed curves are the simulation results.
}
\label{fig4}
\end{figure}

Here, the background counts are measured at the signal and idler channels by blocking the input photons and then subtracted from the raw data when calculating the splitting ratio. See Supplementary Information, Sec.~II, for details on the measurement of the splitting ratio. The solid blue and dashed red lines represent the fitted curves of the envelopes, which exhibit a sinusoidal waveform. As illustrated in Fig.~S1, the maximum depletion rates are achieved at pump powers of 10.4 W and 12.7 W for the forward and backward directions, respectively. These discrepant pump powers indicate the pump power loss during the journey through the interferometer, which is approximately 18\% loss (-0.85 dB). 

We observe the maximum translation efficiencies of 85.9$\pm$1.6\% and 81.5$\pm$1.5\% for the forward and backward directions, respectively. See Supplementary Information, Sec.~II, for details on the translation efficiencies. These high efficiencies are attributed to the dispersion property of the NZDSF, which constraints the leakage to the unintended frequency bands and increases the translation efficiencies, given by the narrow phase-matching bandwidth of 1.32 nm~\cite{mckinstrie2005translation}. We expect that higher translation efficiencies are available by engineering the dispersion property of optical fibers. Additionally, In the quantum frequency translation process, the generation of noise photons via spontaneous four-wave mixing~\cite{park2021telecom} and spontaneous Raman scattering is negligible. This is attributed to the separation of more than 270 nm between the BS-FWM pumps and signal/idler photons. See Supplementary Information, Sec. II, for the measured noise counts.

Fig.~\ref{fig4} represents the net visibility of the HOM interference versus the input BS-FWM pump power. Blue squares and red diamonds depict the net visibilities for the forward and backward directions of BS-FWM, respectively. The solid blue and dashed red lines correlate to the numerical simulations using a Green-function method~\cite{mcguinness2011theory, lee2023translation}, modeling the effect of BS-FWM~\cite{dudley2010supercontinuum, AGRAWAL201327} as an input-output relation. See Supplementary Information, Sec.~III.A. for the simulation method and Sec.~IV for the frequency-domain HOM interference without accidental subtraction. The experimental results match well with the simulation results. 

Creating a NOON interferometer requires two 50:50 beam splitters. Due to a NOON-interferometer transmission of 82\% (-0.85 dB) for the BS-FWM pumps, the HOM visibility for forward and backward propagation peaks at different powers of 5.20 W and 6.32 W, respectively, where the splitting ratio is close to 50\%, and their maximum net (raw) visibility is 97.5$\pm$4.2\% (90.5$\pm$3.8\%) and 91.3$\pm$3.4\% (83.9$\pm$3.0\%), respectively, as seen in Fig.~\ref{fig4}. As the 82\% transmission incurred through the NOON interferometer makes it difficult to obtain a 50\% transmittance for both the forward and backward directions, we utilize an input BS-FWM power of 5.76 W, achieving a transmittance of 45\% in the forward direction, leading to a 60\% transmittance in the backward direction, where the corresponding net (raw) visibilities are 93.0$\pm$3.4\% (86.5$\pm$3.2\%) and 85.5$\pm$3.2\% (80.2$\pm$2.9\%), respectively. These unbalanced splitting ratios will lead to non-ideal visibility in the NOON-state interference.

\begin{figure}[t!]
\centering
\includegraphics[width=3.35in]{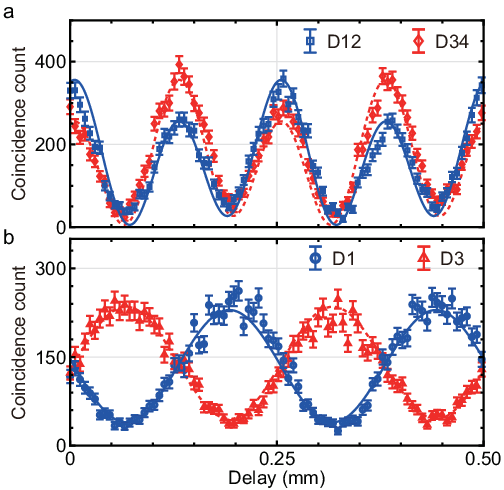}
\caption{\textbf{Frequency-domain NOON-state interference.} 
\textbf{a}, The true coincidence counts versus the temporal delay (DL2) for the NOON-state interference (N = 2). The blue and red points indicate the true coincidence counts of the SNSPDs D1 \& D2 and D3 \& D4, respectively, which enables to post-select two-photon events. The solid blue and dashed red lines are simulation results. \textbf{b}, The true coincidence counts versus the temporal delay (DL2) for the single-photon interference. The blue and red points represent the true coincidence counts of the SNSPDs D1 \& D5 and D3 \& D5. The solid blue and dashed red lines are simulation results. The error bars are calculated assuming Poissonian statistics of the detection.
}
\label{fig5}
\end{figure}

\begin{figure}[t!]
\centering
\includegraphics[width=3.35in]{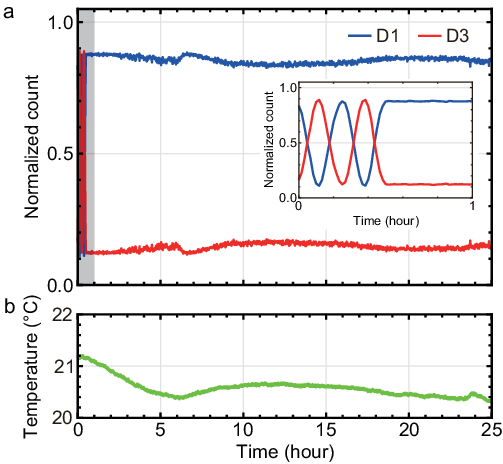}
\caption{\textbf{Stability analysis of the frequency-domain NOON-state interferometer.} 
\textbf{a}, Normalized coincidence counts over 25 hours-time periods using heralded signal-photons. The blue and red curves represent the coincidence detections between the SNSPDs D1 \& D5 and D3 \& D5, respectively. The inset provides a magnified view over a 1-hour period, which is shown as a dashed gray box. \textbf{b}, Temperature measurements from our laboratory, captured during the identical time window of the stability test.
}
\label{fig6}
\end{figure}

The frequency-domain NOON-state interference and single-photon interference patterns are shown in Fig.~\ref{fig5}. The NOON interference is gauged using the experimental setup described in Fig.~\ref{fig2}. Fig.~\ref{fig5}\textbf{a} depicts the NOON-state interference, where the blue squares and red diamonds indicate the true coincidence counts between the SNSPDs D1 \& D2 and D3 \& D4, respectively. Each point is accumulated for 60 seconds. The solid blue and dashed red lines are the simulation results, which align closely with the experimental findings. See Supplementary Information, Sec.~III.B, for the simulation method of the frequency-domain NOON-state interference. Notably, the HOM interference in Fig.~\ref{fig4} and the two-fold enhanced quantum resolution in Fig.~\ref{fig5} are the signature of the generation of the 2002 state, $\ket{\psi}= \frac{1}{\sqrt{2}} (\ket{2}_|s|\ket{0}_|i| + \ket{0}_|s|\ket{2}_|i|)$, commonly referred to as the photon number path entangled state. Our results demonstrate the achievement of super-resolution within this framework. However, it is important to note that super-sensitivity was not observed, which is attributable to system losses. See Supplementary Information, Sec.~V for detailed calculations about Fisher Information and super-sensitivity.

In comparison, Fig.~\ref{fig5}\textbf{b} presents the single-photon interference, which has been implemented using a slightly modified version of the setup depicted in Fig.~\ref{fig2}. Rather than deploying both signal and idler photons into the interferometer, the idler photon is employed as a trigger photon using a heralding detector D5, which is not shown in Fig.~\ref{fig2}, and the heralded signal photon is directed toward the interferometer. Two beam splitters preceding the detectors D1 and D3 are removed to enhance the count rates. The blue circles and red triangles indicate the true coincidence counts between the SNSPD D1 \& D5 and D3 \& D5, respectively. Each data point is acquired for one second. The solid blue and dashed red lines are the theoretical results, which correlate closely with the experimental data. See Supplementary Information, Sec.~III.C for the simulation method of the single-photon interference and Sec.~VI for the measurement results without subtracting the accidental counts. By fitting the oscillations with a sinusoidal function, we extracted the oscillation periods as 0.125±0.001 mm for the NOON state and 0.248±0.001 mm for the single-photon state. Theoretically, these periods are calculated to be 0.125 mm and 0.250 mm, respectively, derived using the relation $Period = c/(2N \Delta f)$. The factor of two in this formula arises from the double pass of the delay line in our experiment. The experimentally determined oscillation periods align well with these theoretical expectations. While our current measurements focus on the 2002-interference pattern, our experiment setup is adaptable for the measurement of the 4004-interference pattern via post-selection~\cite{nagata2007beating, shin2013enhancing}.

Furthermore, our interferometer shows outstanding stability due to the use of a single-mode spatial path. Two frequency modes propagate a single-mode fiber, and this guarantees that the phase shifts of the two frequency modes will be almost identical, effectively neutralizing the phase difference between them. Such attributes enable the frequency-domain interferometer with exceptional stability. Fig.~\ref{fig6}\textbf{a} illustrates the experimental results of a single-photon interferometer stability test conducted over a 25-hour period. The blue and red curves represent the normalized coincidence counts between SNSPDs D1 \& D5 and D3 \& D5, respectively, where D5 is the heralding detector. Each data point is accumulated in one-minute intervals. For the first 0.5 hours, we introduce a two-cycle of phase change by tuning the temporal delay line (DL2). Then, for the remainder of the test duration, this path-length variation is halted. The inset of Fig.~\ref{fig6}\textbf{a} represents an expanded view of Fig.~\ref{fig6}\textbf{a}, captured during the first hour of measurement. The measured interference patterns display remarkable stability. Of particular note, this experiment was performed on an unfloated optical table with a hundred meters of optical fibers. Moreover, the setup was located under an air conditioner without any active temperature/phase control or wind protection. The expected temperature stability of our interferometer, including the double-pass propagation of the 100-m NZDSF and additional meters of single-mode fibers, is roughly $\sim$ 0.4 $\pi /\degree \mathrm{C}$, which is calculated as 1.4 $\times$ $10^{-3}$ ps/nm/km/\textdegree C $\times$ 3.2 nm $\times$ 0.2 km $\times$ 2$\pi$c/1270 nm. This calculation is based on a chromatic dispersion thermal coefficient of 1.4 $\times$ $10^{-3}$ ps/nm/km/\textdegree C, derived from measurements on a single-mode fiber at 1550 nm~\cite{andre2005chromatic}. Fig.~\ref{fig6}\textbf{b} represents the measured temperature stability within our laboratory during the same time frame as our stability test. The average temperature is 20.6$\pm$0.2\textdegree C throughout this 25-hour period. Under these conditions, the frequency-domain interferometer exhibits unprecedented stability.

In conclusion, we demonstrated the pioneering NOON-state interference in the frequency domain by employing frequency beam splitters and non-degenerate photon pairs. The NOON state's relative phase is controlled by a delay line, resulting in two-photon interference. The resulting observations include the two-photon frequency bunching and anti-bunching effects depending on the relative phase between the two frequency modes, two-fold enhancement in the oscillation period compared to that of a single-photon interference, and unprecedented stability of the interferometer. 

\section*{Discussion}
The NOON state exhibits a non-sinusoidal interference pattern, while the single-photon state shows a sinusoidal pattern. In addition, both patterns exhibit non-ideal visibility. The following three factors contribute to this non-sinusoidal pattern and non-ideal visibility: 1) asymmetricity in the joint-spectral intensity, 2) dispersion effect of the NZDSF causing a walk-off between the two photons, and 3) Unbalanced BS-FWM pump powers for forward and backward propagation. These three effects can be mitigated by the following strategies: 1) Reducing the length of the nonlinear medium (SMF) for the photon-pair generation to relax the phase-matching condition~\cite{park2021telecom}. 2) Placing a dispersion compensation component to avoid the walk-off effect, such as a pair of gratings or an optical Bragg filter. 3) Amplifying the BS-FWM pumps to compensate for the NOON interferometer insertion loss. See Supplementary Information, Sec.~III, for details on these three factors contributing to non-ideal visibility.

In this study, a high pump power of up to 15 W was utilized to compensate for the low nonlinearity coefficient of the optical fiber. Despite this, the generation of noise photons remained negligibly low, posing no significant issue. In addition, this high pump power requirement significantly decreases when utilizing on-chip platforms. A demonstration in a silicon waveguide reveals an efficiency of 12\% with the BS-FWM pump power of 1.6 mW~\cite{bell2016frequency}. Additionally, silicon nitride resonators report an efficiency of 60\% with a pump power of 60 mW~\cite{li2016efficient}. This subject is fascinating but not within our current research scope. It is, however, a prime candidate for our future research projects.

With its scalability to high-dimensional multi-frequency states and the potential for miniaturization through single-mode propagation, quantum information processing in the frequency domain has attracted significant attention, especially in the context of quantum communications~\cite{lu2023frequency}. As shown in this study, frequency-domain quantum information processing highlights the compatibility with optical fiber networks, in addition to demonstrating the remarkable stability of our interferometer, a crucial resource in both fundamental research and application~\cite{lu2023recent, yang2021angular, geng2022surface}, including quantum communications~\cite{zhang2022realization} and linear quantum computing~\cite{dellios2023simulating}.

Owing to the excellent stability and scalability of our quantum states, our work can be extended to implement high-dimensional quantum key distribution (QKD) through deployed fiber networks. This method enhances information efficiency and may extend the maximum transmission distance compared to conventional two-dimensional QKD systems~\cite{ding2017high}. Notably, high-dimensional states in the frequency domain can be readily generated by inserting optical filters or a Fabry-Perot cavity after the creation of photon pairs~\cite{lu2023frequency}. Quantum simulators can be investigated within the frequency domain, including quantum random walks~\cite{tang2018experimental} and Boson sampling~\cite{dellios2023simulating}. By varying the number of BS-FWM pumps and adjusting the relative phase and power between them, we can explore a variety of quantum-circuit configurations and interesting phenomena, including non-local hopping~\cite{bell2017spectral, wang2020multidimensional}. These features may offer insights into areas not readily accessible through conventional path-mode schematics. Finally, our NOON-state interferometer, exploiting the quantum interference across different frequencies, is capable of measuring wavelength-dependent phase shifts. This feature holds the potential to implement quantum quantitative phase spectroscopy, which investigates the properties of organisms or cells, especially those with extremely low damage thresholds~\cite{clemmen2016ramsey, rinehart2012quantitative}.

\bibliographystyle{naturemag}
\bibliography{reference}

\clearpage

\section*{acknowledgements}
\noindent
This work was supported by the National Research
Foundation of Korea (NRF-2019M3E4A1079780) and
IITP (Institute for Information \& Communications Technology Planning \& Evaluation) grant funded by the Korea government (MSIT) (IITP-2022-RS-2022-00164799 and No.2020-0-00947).

\section*{Author contributions}
\noindent
D.L. and H.S. proposed the idea and designed the experiment. D.L., S.P., and J.K. calculated the theoretical results. D.L. carried out the experimental and simulation works with support from W.S.. All authors contributed to writing the manuscript and participated in the data analysis and discussions. H.S. supervised the project.

\section*{Conflict of Interest statement}
\noindent
The authors declare no competing interests.

\section*{Methods}
\noindent
\textbf{Preparation of the photon pair.} 
A non-degenerate photon pair is created via spontaneous four-wave mixing within a 200-m-long single-mode fiber (Corning, SMF-28), mediated by a pump laser under the relaxed phase-matching condition~\cite{park2021telecom}. The pump laser features a central wavelength of 1269.50 nm, a duration of 100 ps, and a peak power of about 1 W. The laser operates at a repetition rate of 20 MHz, electronically synchronized with the BS-FWM pump lasers. We employ two strategies to reduce the noise photons resulting from spontaneous Raman scattering. First, we cool the single-mode fiber by submerging it in liquid nitrogen, thereby reducing the population of excited state phonons. Second, we attach a polarizer to the output of the single-mode fiber and align the polarization direction of the pump to that of the polarizer. This is because the polarization state of the photon pair is parallel to that of the pump laser, while noise photons are unpolarized. The pump, signal, and idler photons are separated by bandpass filters with a bandwidth of 0.7 nm and centered at 1269.50 nm, 1267.89 nm, and 1271.11 nm, respectively. The setup ensures a pump rejection ratio exceeding 120 dB. The resultant coincidence-to-accidental ratio for the photon pair is 22.2$\pm$1.1.

\end{document}


\title{Supplementary Information for NOON-state interference in the frequency domain}

\author{Dongjin Lee}
\affiliation{Department of Physics, Pohang University of Science and Technology (POSTECH), Pohang, 37673, South Korea}

\author{Woncheol Shin}
\affiliation{Department of Physics, Pohang University of Science and Technology (POSTECH), Pohang, 37673, South Korea}

\author{Sebae Park}
\affiliation{Department of Physics, Pohang University of Science and Technology (POSTECH), Pohang, 37673, South Korea}

\author{Junyeop Kim}
\affiliation{Department of Physics, Pohang University of Science and Technology (POSTECH), Pohang, 37673, South Korea}

\author{Heedeuk Shin}
\email{heedeukshin@postech.ac.kr}
\affiliation{Department of Physics, Pohang University of Science and Technology (POSTECH), Pohang, 37673, South Korea}

\date{\today}
\maketitle

\section{Background theory}
In this section, we will derive analytic solutions of the Hong-Ou-Mandel (HOM) effect, NOON-state interference, and single-photon interference in the frequency domain. We consider the single-mode inputs to derive the analytic solutions. The solutions for these effects with pulsed mode inputs will be calculated in Sec.~\ref{Simulation_method}.

\subsection{Frequency-domain HOM interference}
As we discussed in the main text, the evolution of the annihilation mode operators via Bragg scattering four-wave mixing (BS-FWM) is described by the following equation~\cite{mckinstrie2005translation, mcguinness2010quantum, clemmen2016ramsey},

\begin{equation}
\label{eqB1}
\begin{bmatrix}
\hat{a}_|s,out| \\
\hat{a}_|i,out|
\end{bmatrix} = 
\begin{bmatrix}
\cos(gL) & ie^{i\phi} \sin(gL) \\
ie^{-i\phi} \sin(gL) & \cos(gL) 
\end{bmatrix} 
\begin{bmatrix}
\hat{a}_|s,in| \\
\hat{a}_|i,in|
\end{bmatrix}.
\end{equation}

\noindent 
Here, the coefficient $g$ is defined by $g \equiv \gamma P$, where $\gamma$ and $P$ denote the nonlinear coefficient of the nonlinear medium and the total pump power of BS-FWM, respectively. The term $\phi$ denotes the phase difference between the two pump beams. This equation resembles the transformations in beam splitters, allowing us to interpret BS-FWM as a frequency beam splitter~\cite{clemmen2016ramsey, kobayashi2016frequency}. While the coefficient $\phi$ appears unusual, we will confirm that this coefficient does not affect the measured outcomes. For ease of calculation, we rewrite Eq.~(\ref{eqB1}) by representing the input-creation operators in terms of the output-creation operators, 

\begin{equation}
\label{eqB2}
\begin{bmatrix}
\hat{a}_|s,in|^\dagger \\
\hat{a}_|i,in|^\dagger
\end{bmatrix} = 
\begin{bmatrix}
\cos(gL) & ie^{-i\phi} \sin(gL) \\
ie^{i\phi} \sin(gL) & \cos(gL) 
\end{bmatrix} 
\begin{bmatrix}
\hat{a}_|s,out|^\dagger \\
\hat{a}_|i,out|^\dagger
\end{bmatrix}.
\end{equation}

When inserting two single photons of different frequencies (signal and idler) into a frequency beam splitter, the state is transformed as

\begin{equation}
\label{eqB3}
\begin{aligned}
    &\hat{a}_|s,in|^\dagger \hat{a}_|i,in|^\dagger \ket{0} = [\cos(gL) \hat{a}_|s,out|^\dagger + ie^{-i\phi} \sin(gL) \hat{a}_|i,out|^\dagger] [ie^{i\phi} \sin(gL) \hat{a}_|s,out|^\dagger + \cos(gL) \hat{a}_|i,out|^\dagger] \ket{0} \\
    &= ie^{i\phi} \cos(gL) \sin(gL) \hat{a}_|s,out|^\dagger \hat{a}_|s,out|^\dagger \ket{0}+ ie^{-i\phi} \cos(gL) \sin(gL) \hat{a}_|i,out|^\dagger \hat{a}_|i,out|^\dagger \ket{0} + \left[ \cos^2(gL) - \sin^2(gL) \right] \hat{a}_|s,out|^\dagger \hat{a}_|i,out|^\dagger \ket{0}. \\
\end{aligned}
\end{equation}

\noindent
At the splitting ratio of 50\% ($gL = \pi/4$), the HOM effect in the frequency domain can be observed,  as indicated by
\begin{equation}
\label{eqB4}
\begin{aligned}
    \hat{a}_|s,in|^\dagger \hat{a}_|i,in|^\dagger \ket{0} &= \frac{i}{2} e^{i\phi} \hat{a}_|s,out|^\dagger \hat{a}_|s,out|^\dagger \ket{0} + \frac{i}{2} e^{-i\phi} \hat{a}_|i,out|^\dagger \hat{a}_|i,out|^\dagger \ket{0} \\ &= \frac{i}{\sqrt{2}} e^{i\phi} \ket{2}_|s,out| \ket{0}_|i,out| + \frac{i}{\sqrt{2}} e^{-i\phi} \ket{0}_|s,out| \ket{2}_|i,out|,
\end{aligned}
\end{equation}

\noindent
where $\ket{N}_|m,out|$ ($\rm m = s, i$) represent the Fock states (N-photon-number states) at the output signal and idler modes, respectively.

\subsection{Frequency-domain NOON-state interference}
The function of the interferometer in Fig.~1\textbf{d} can be represented by the following matrix multiplications, 

\begin{equation}
\label{eqB5}
\begin{bmatrix}
\hat{a}_|s,in|^\dagger \\
\hat{a}_|i,in|^\dagger
\end{bmatrix} = \frac{1}{2}
\begin{bmatrix}
1 & ie^{-i\phi}  \\
ie^{i\phi} & 1 
\end{bmatrix} 
\begin{bmatrix}
\exp[-\frac{i 2\pi f_|s| L}{c}] & 0 \\
0 & \exp[-\frac{i 2\pi f_|i| L}{c}]
\end{bmatrix} 
\begin{bmatrix}
1 & ie^{-i\phi}  \\
ie^{i\phi} & 1 
\end{bmatrix} 
\begin{bmatrix}
\hat{a}_|s,out|^\dagger \\
\hat{a}_|i,out|^\dagger
\end{bmatrix},
\end{equation}

\noindent
where the first and third terms represent the frequency beam splitters, and the second term indicates the phase term, introduced by the distance between the two splitters. We assumed that the splitting ratios of the beam splitters are 50\% ($gL = \pi/4$). The terms $f_|s|$ and $f_|i|$ are the frequencies of the signal and idler modes, respectively, and $L$ is the distance between the two splitters. The resultant result from Eq.~(\ref{eqB5}) is expressed as

\begin{equation}
\label{eqB6}
\begin{aligned}
\begin{bmatrix}
\hat{a}_|s,in|^\dagger \\
\hat{a}_|i,in|^\dagger
\end{bmatrix} &= \frac{1}{2} \exp \left[-\frac{i 2\pi f_|s| L}{c} \right]
\begin{bmatrix}
1 & ie^{-i\phi}  \\
ie^{i\phi} & 1 
\end{bmatrix} 
\begin{bmatrix}
1 & 0 \\
0 & e^{i\Delta \phi}
\end{bmatrix} 
\begin{bmatrix}
1 & ie^{-i\phi}  \\
ie^{i\phi} & 1 
\end{bmatrix} 
\begin{bmatrix}
\hat{a}_|s,out|^\dagger \\
\hat{a}_|i,out|^\dagger
\end{bmatrix} \\
&= \frac{1}{2} \exp \left[-\frac{i 2\pi f_|s| L}{c} \right]
\begin{bmatrix}
1 - e^{i\Delta \phi} & ie^{-i\phi} (1 + e^{i\Delta \phi}) \\
ie^{i\phi} (1 + e^{i\Delta \phi}) & -1 + e^{i\Delta \phi}
\end{bmatrix}
\begin{bmatrix}
\hat{a}_|s,out|^\dagger \\
\hat{a}_|i,out|^\dagger
\end{bmatrix},
\end{aligned}
\end{equation}

\noindent
where $\Delta \phi = 2\pi \Delta f L/c$ and $\Delta f = f_|s| - f_|i|$. We will ignore the global phase factor for further calculations, as it does not influence the measurement results. Inserting two single photons (signal and idler photons) into the interferometer, the output state is given by

\begin{equation}
\label{eqB7}
\begin{aligned}
    \hat{a}_|s,in|^\dagger \hat{a}_|i,in|^\dagger \ket{0} &= \frac{i}{4} e^{i\phi} [1 - e^{2i\Delta \phi}] \hat{a}_|s,out|^\dagger \hat{a}_|s,out|^\dagger \ket{0} - \frac{i}{4} e^{-i\phi} [1 - e^{2i\Delta \phi}] \hat{a}_|i,out|^\dagger \hat{a}_|i,out|^\dagger \ket{0} - \frac{1}{2} [1 + e^{2i \Delta \phi}] \hat{a}_|s,out|^\dagger \hat{a}_|i,out|^\dagger \ket{0} \\
    &= \frac{1}{\sqrt{2}} e^{i(\Delta \phi + \phi)} \sin(\Delta \phi) \ket{2}_|s,out| \ket{0}_|i,out| - \frac{1}{\sqrt{2}} e^{i(\Delta \phi - \phi)} \sin(\Delta \phi) \ket{0}_|s,out| \ket{2}_|i,out| - e^{i\Delta \phi} \cos(\Delta \phi) \ket{1}_|s,out| \ket{1}_|i,out|.
\end{aligned}
\end{equation}

\noindent
The oscillation period for the two-photon detection for the NOON-state interference (N = 2) is $c/(2\Delta f)$.

\subsection{Frequency-domain single-photon interference}
The single-photon interference can be calculated with Eq.~(\ref{eqB6}). When a signal photon is injected into the interferometer, the output state is expressed as 

\begin{equation}
\label{eqB8}
\begin{aligned}
    \hat{a}_|s,in|^\dagger \ket{0} &= \frac{1}{2} [1 - e^{i\Delta \phi}] \hat{a}_|s,out|^\dagger \ket{0} + \frac{i}{2} e^{-i\phi} [1 + e^{i\Delta \phi}] \hat{a}_|i,out|^\dagger \ket{0} \\
    &= -ie^{i \Delta \phi/2} \sin(\Delta \phi /2) \ket{1}_|s,out| \ket{0}_|i,out| + ie^{i \Delta \phi/2 - i\phi} \cos(\Delta \phi/2) \ket{0}_|s,out| \ket{1}_|i,out|,
\end{aligned}
\end{equation}

\noindent
where the oscillation period is $c/(\Delta f)$, twice as long as that for the NOON-state interference (N = 2).\\

In conclusion, we have calculated the Hong-Ou-Mandel effect, NOON-state interference, and single-photon interference in the frequency domain. Despite the presence of the coefficient $\phi$ in Eq.~(\ref{eqB1}), which differs from typical beam splitter operators, it does not affect the outcome of the measurements.

\clearpage
\section{Detailed information about the single-photon translation}

\begin{figure*}[b!]
\centering
\includegraphics[width=3.35in]{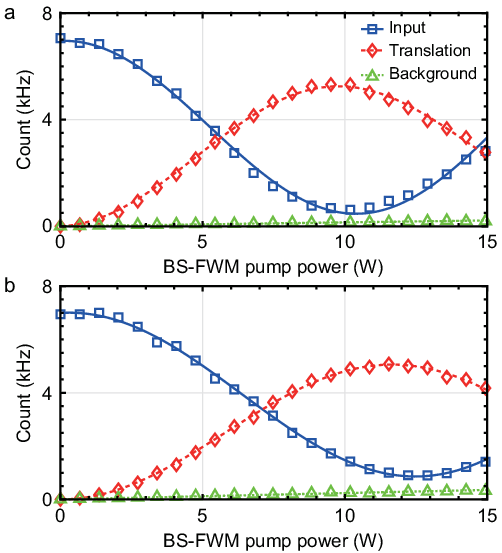}
\caption{\textbf{Single-photon translation of the BS-FWM effect.} The single-photon translation is measured by blocking the idler photons and increasing the delay line (DL2) to allow the quantum frequency translation for the single direction. The blue squares and red diamonds indicate the measured counts at the signal (input) and idler (translation) wavelength, respectively. The background counts (green triangle) represent the measured counts at the idler wavelength while the signal photons are blocked. The lines represent the best-fit curves. \textbf{a}, the forward direction. \textbf{b}, the backward direction. 
}
\label{sfig1_1}
\end{figure*}

The single-photon translation of the BS-FWM effect for the forward direction (a) and the backward direction (b) is described in Fig.~\ref{sfig1_1}. The experimental setup in Fig.~2 is slightly modified to measure the translation. The signal photons are used as the input photons, while the idler photons are blocked. The delay line (DL2) is increased as much as 4.5 ns to activate the BS-FWM effect only for the single direction. The output of each channel is measured by an SNSPD without the beam splitter to increase the count rate. The blue squares and red diamonds are the measured counts at the signal (input) and idler (translation) wavelength, respectively. To measure the noise of the BS-FWM effect, the background counts are measured at the signal and idler wavelength while the input photons are blocked. As the background counts at each wavelength are similar, we represent the background counts (green triangles) at only the idler (translation) wavelength. Notably, the background counts can be reduced by cooling the DSF with liquid nitrogen, decreasing noise photons originating from spontaneous Raman scattering~\cite{clemmen2016ramsey, joshi2020frequency}. The solid blue, dashed red, and dotted green lines represent the best-fit curves of the input, translation, and background counts where the fit functions are given by $a \cos^2(b p) + c p$, $a \sin^2(b p) + c p$, and $a p$, respectively. $a, b, c$ are the fit parameters and $p$ is the total power of the BS-FWM. 

We calculate the calibrated counts by compensating the background counts and the relative efficiency between the channels. The relative efficiency of the idler channel is 85\% (-0.7 dB) compared to that of the signal channel. With the calibrated counts, we achieve the efficiencies and the fit depletion rates of the BS-FWM effect. The efficiencies are 85.9$\pm$1.6\% and 81.5$\pm$1.5\%, and the fit depletion rates are 95.2\% and 92.6\% for the forward and the backward directions, respectively. The maximum depletion occurs at the different input powers for the forward (10.4 W) and backward directions (12.7 W), respectively, as the NOON-state interferometer introduces the loss to the BS-FWM pump. We attribute the discrepancy between the translation efficiencies to the imperfection in polarization compensation. We use two Faraday mirrors, denoted as FM1 and FM2 in Fig. 2, with the optimized wavelengths of 1310 nm and 1550 nm. While the BS-FWM pumps are within FM1’s operation wavelength range at near 1550 nm, but the signal photon is far from FM2’s optimized wavelength of 1310 nm. This leads to imperfect polarization compensation, which results in slightly lower translation efficiency for the backward pumping. 
We expect that the translation efficiency and splitting ratio are identical for the input idler photons. Theoretically, we can identify from Eq. (1) and (2) in the main text that the operator forms for the signal and idler modes are equivalent. Experimentally, Clark et al.~\cite{clark2013high} demonstrate equivalent translation efficiencies of 99.1\%$\pm$4.9\% and 98.0$\pm$5.0\% for the input idler and signal photons, respectively.

\clearpage
\section{Simulation method} \label{Simulation_method}
\subsection{Frequency-domain HOM interference}
\begin{figure}[h]
\centering
\includegraphics[width=6.7in]{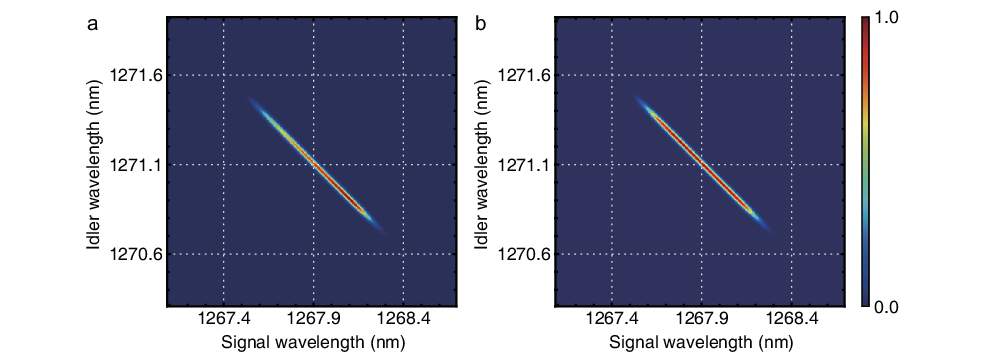}
\caption{\textbf{Simulated joint spectral intensity of the photon pair.}
\textbf{a}, The simulated conditions are identical to our experimental conditions, where the length of the optical fiber is 200 m. \textbf{b}, The length of the optical fiber is shortened to 50 m, but all other conditions remain consistent.
}
\label{sfig2_1}
\end{figure}

We first calculate the joint spectral intensity of the photon-pair generation under our experimental conditions as described in Fig.~\ref{sfig2_1}\textbf{a}. One of the reasons for having non-unity visibility in Fig.~4, even in theory, is its asymmetrical joint-spectral intensity shape of the photon-pair source under our experimental conditions. See the last of the sections for more details.
Note that the asymmetricity in the joint spectral intensity can be resolved by reducing the length of the single-mode fiber (SMF) to relax the phase matching condition~\cite{park2021telecom}. As seen from Fig.~\ref{sfig2_1}\textbf{b}, we barely observe the asymmetricity by reducing the length of the SMF from 200 m to 50 m. 

For convenience in the simulation, Schmidt decomposition is applied to the joint spectral intensity, which represents the correlated function in terms of a linear combination, 

\begin{equation}
\label{eqS1}
    \ket{\psi_|in|} = \int d\omega_|s| d\omega_|i| F(\omega_|s|, \omega_|i|) \hat{a}^\dagger_|s|(\omega_|s|) \hat{a}^\dagger_|i|(\omega_|i|) \ket{0} = \sum_|k| r_|k| \int d\omega_|s| F_|k|(\omega_|s|) \hat{a}^\dagger_|s|(\omega_|s|) \int d\omega_|i| G_|k|(\omega_|i|) \hat{a}^\dagger_|i|(\omega_|i|) \ket{0},
\end{equation}

\noindent
where $F(\omega_|s|, \omega_|i|)$ represents the joint spectral amplitude and $F_|k|(\omega)$ and $G_|k|(\omega)$ are sets of orthonormal basis, known as Schmidt modes. Schmidt amplitude $r_|k|$ (real number) weights the intensity of each set where $\sum_|k| r_|k|^2 = 1$. $\hat{a}^\dagger_|s|(\omega)$ is a creation operator at the signal mode with a frequency of $\omega + \omega_|s0|$, and $\hat{a}^\dagger_|i|(\omega)$ is a creation operator at the idler mode with a frequency of $\omega + \omega_|i0|$. $\omega_|s0|$ and $\omega_|i0|$ are the central frequencies of bandpass filters (FWHM: 0.7 nm) which are used to refine the joint spectral intensity, where $\omega_|s0| = 2\pi \, \times$ 236.45 THz (1267.89 nm) and $\omega_|i0| = 2\pi \, \times$ 235.85 THz (1271.11 nm), respectively. \\

The Bragg scattering four-wave mixing (BS-FWM) effect is numerically implemented with a Green-function method~\cite{mcguinness2011theory}, which represents the effect in terms of an input-to-output relation. To construct the Green function, we use Hermite-Gaussian (HG) functions as an orthogonal set of the input amplitudes and calculate the corresponding output amplitudes with coupled equations governing the BS-FWM effect. The coupled equations are derived from the nonlinear Schr\"{o}dinger equation (NLSE)~\cite{mcguinness2011theory, dudley2010supercontinuum, AGRAWAL201327}. More details about the construction of the Green function are described in our previous paper~\cite{lee2023translation}. As the length of the SMF is different (150 m) in the previous paper, we use an optimal characteristic length of the HG functions as 29 ps instead of 28 ps to calculate the Green function. The constructed Green function describes the transformation of the creation operators at the signal and idler modes via BS-FWM as follows, 

\begin{equation}
\label{eqS2}
    \begin{aligned}
        \hat{a}^\dagger_|s|(\omega) \rightarrow & \int d\omega' G_|ss|(\omega, \omega') \hat{a}^\dagger_|s|(\omega') + G_|si|(\omega, \omega') \hat{a}^\dagger_|i|(\omega'), \\
        \hat{a}^\dagger_|i|(\omega) \rightarrow & \int d\omega' G_|is|(\omega, \omega') \hat{a}^\dagger_|s|(\omega') + G_|ii|(\omega, \omega') \hat{a}^\dagger_|i|(\omega'),
    \end{aligned}
\end{equation}

\noindent
where $G_|ij|(\omega, \omega’)$ ($\rm{i,j = s,i}$) is the Green function which indicates the evolution amplitude describing the frequency component ($\omega$) at the input mode $\rm{i}$ is translated to the other component ($\omega’$) at the output mode $\rm{j}$. \\

Now, we are ready to calculate the frequency-domain HOM interference. The delay line (DL1) introduces a temporal delay ($\Delta t_|1|$) to the idler photon, transforming equation (\ref{eqS1}) as below

\begin{equation}
\label{eqS3}
    \ket{\psi_|in|} \rightarrow \ket{\psi_|in,1|} = \sum_|k| r_|k| \int d\omega_|s| F_|k|(\omega_|s|) \hat{a}^\dagger_|s|(\omega_|s|) \int d\omega_|i| G_{\rm{\Delta t_1, k}}(\omega_|i|) \hat{a}^\dagger_|i|(\omega_|i|) \ket{0},
\end{equation}

\noindent
where $G_{\rm{\Delta t_1, k}}(\omega) = G_|k|(\omega) \exp[i(\omega_|i0| + \omega) \Delta t_1]$. The BS-FWM process, implementing the role of the frequency beam splitter, evolves the state as

\begin{equation}
\label{eqS4}
    \ket{\psi_|out,1|} = \sum_|k| r_|k| \left( \int d\omega_|s| F_|ss,k|(\omega_|s|) \hat{a}^\dagger_|s|(\omega_|s|) + F_|si,k|(\omega_|s|) \hat{a}^\dagger_|i|(\omega_|s|) \right) 
    \times \left( \int d\omega_|i| G_{\rm{is,\Delta t_1,k}}(\omega_|i|) \hat{a}^\dagger_|s|(\omega_|i|) + G_{\rm{ii,\Delta t_1,k}}(\omega_|i|) \hat{a}^\dagger_|i|(\omega_|i|) \right) \ket{0},
\end{equation}

\noindent
where the amplitudes $F_|xy,k|(\omega)$ and $G_{\rm{xy,\Delta t_1,k}}(\omega)$ ($\rm{x,y = s, i}$) indicate  $F_|xy,k|(\omega) = \int d\omega' F_|k|(\omega') G_|xy|(\omega',\omega)$ and $G_{\rm{xy,\Delta t_1,k}}(\omega) = \int d\omega' G_{\rm{\Delta t_1,k}}(\omega') G_|xy|(\omega',\omega)$, respectively. Each term ($F_|xy,k|(\omega)$ and $G_{\rm{xy,\Delta t_1,k}}(\omega)$) indicates the output amplitude in the mode $\rm{y}$ for the given input amplitude ($F_|k|(\omega)$ and $ G_{\rm{\Delta t_1,k}}(\omega)$) in the mode $\rm{x}$ via the BS-FWM effect. \\

Lastly, we consider a coincidence detection between the idler and signal modes for the final state ($\ket{\psi_|out,1|}$). For convenience, we rewrite the final state in equation (\ref{eqS4}) as

\begin{equation}
\label{eqS5}
    \ket{\psi_|out|} = \sum_|k| r_|k| \left( \int d\omega_s A_|k|(\omega_|s|) \hat{a}_s^\dagger(\omega_|s|) + B_|k|(\omega_|s|) \hat{a}^\dagger_|i|(\omega_|s|) \right) \times
    \left( \int d\omega_|i| C_|k|(\omega_|i|) \hat{a}^\dagger_|s|(\omega_|i|) + D_|k|(\omega_|i|) \hat{a}^\dagger_|i|(\omega_|i|) \right) \ket{0}.
\end{equation}

\noindent
The two-photon detection of the final state at time $t_1$ in the signal mode and at time $t_2$ in the idler mode is described by 

\begin{equation}
\label{eqS6}
\hat{E}_|s|(t_1) \hat{E}_|i|(t_2) \ket{\psi_|out|} = \sum_|k| r_|k| \int d\omega_|1| d\omega_|2| [A_|k|(\omega_|1|) D_|k|(\omega_|2|) + C_|k|(\omega_|1|) B_|k|(\omega_|2|)] \exp[-i \omega_|1| t_|1| - i \omega_|2| t_|2|] \ket{0},
\end{equation}

\noindent
where the electric-field annihilation operator $E_|x|(t)$ is $E_|x|(t) = \int d\omega \hat{a}_x(\omega) e^{-i \omega t}$. The probability density of the two-photon detection is expressed as 

\begin{equation}
\label{eqS7}
    \begin{aligned}
        ||\hat{E}_|s|(t_1) \hat{E}_|i|(t_2) \ket{\psi_|out|}|^2 =  \sum_|k, k'| r_|k| r^*_|k'| \int d\omega'_|1| d\omega'_|2| d\omega_|1| d\omega_|2| [A^*_|k'|(\omega'_|1|) D^*_|k'|(\omega'_|2|) + C^*_|k'|(\omega'_|1|) B^*_|k'|(\omega'_|2|)] \\
        \times [A_|k|(\omega_|1|) D_|k|(\omega_|2|) + C_|k|(\omega_|1|) B_|k|(\omega_|2|)] \exp[i(\omega_|1|' - \omega_|1|)t_|1| + i(\omega_|2|' - \omega_|2|)t_|2|].
    \end{aligned}
\end{equation}

\noindent
The measured coincidence count is proportional to the time-averaged probability as the coherence time of the single photons ($\sim$ ps) is much finer than the temporal resolution of the single-photon detection ($\sim$ 100 ps). The time-averaged probability of the two-photon detection is given by 
\begin{equation}
\label{eqS8}
    \begin{aligned}
        \int dt_|1| dt_|2| ||\hat{E}_|s|(t_1) \hat{E}_|i|(t_2) \ket{\psi_|out|}|^2 = \sum_|k,k'| r_|k| r^*_|k'| \int d\omega_|1| d\omega_|2| & [A^*_|k'|(\omega_|1|) D^*_|k'|(\omega_|2|) + C^*_|k'|(\omega_|1|) B^*_|k'|(\omega_|2|)] \\
        \times & [A_|k|(\omega_|1|) D_|k|(\omega_|2|) + C_|k|(\omega_|1|) B_|k|(\omega_|2|)].
    \end{aligned}
\end{equation}

\noindent
By substituting the parameters in equation (\ref{eqS8}) as $A_|k|(\omega) = F_|ss,k|(\omega)$, $B_|k|(\omega) = F_|si,k|(\omega)$, $C_|k|(\omega) = G_{\rm{is, \Delta t_1, k}}(\omega)$, and $D_|k|(\omega) = G_{\rm{ii, \Delta t_1, k}}(\omega)$, we can calculate the frequency-domain HOM effect. \\

Similarly, we are able to calculate the time-averaged probability of the two-photon detection either in the idler mode or the signal mode. The time-averaged probability in the signal mode is described as

\begin{equation}
\label{eqS9}
    \begin{aligned}
    \int dt_|1| dt_|2| ||\hat{E}_|s|(t_1) \hat{E}_|s|(t_2) \ket{\psi_|out|}|^2  = \sum_|k,k'| r_|k| r^*_|k'| \int d\omega_|1| d\omega_|2| & [A^*_|k'|(\omega_|1|) C^*_|k'|(\omega_|2|) + C^*_|k'|(\omega_|1|) A^*_|k'|(\omega_|2|)] \\
    \times & [A_|k|(\omega_|1|) C_|k|(\omega_|2|) + C_|k|(\omega_|1|) A_|k|(\omega_|2|)],
    \end{aligned}
\end{equation}

and the time-averaged probability in the idler mode is represented as

\begin{equation}
\label{eqS10}
    \begin{aligned}
    \int dt_|1| dt_|2| ||\hat{E}_|i|(t_1) \hat{E}_|i|(t_2) \ket{\psi_|out|}|^2  = \sum_|k,k'| r_|k| r^*_|k'| \int d\omega_|1| d\omega_|2| & [B^*_|k'|(\omega_|1|) D^*_|k'|(\omega_|2|) + D^*_|k'|(\omega_|1|) B^*_|k'|(\omega_|2|)] \\
    \times & [B_|k|(\omega_|1|) D_|k|(\omega_|2|) + D_|k|(\omega_|1|) B_|k|(\omega_|2|)].
    \end{aligned}
\end{equation}

\begin{figure}[h]
\centering
\includegraphics[width=6.7in]{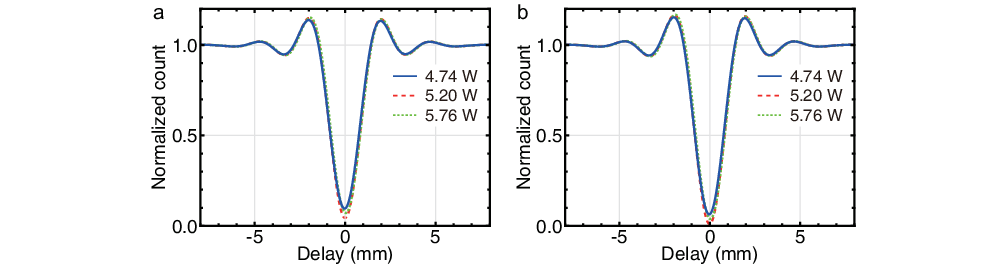}
\caption{\textbf{Simulated frequency-domain HOM effect.}
\textbf{a}, HOM-dip patterns corresponding to various BS-FWM pump powers, where the length of the optical fiber is set to 200 m. \textbf{b}, HOM-dip patterns when the length of the optical fiber is set to 50 m. The BS-FWM pump powers for the blue solid, red dashed, and green dotted curves are 4.74 W, 5.20 W, and 5.76 W, respectively.
}
\label{sfig2_2}
\end{figure}

Fig.~\ref{sfig2_2} represents the simulated frequency-domain HOM patterns for various BS-FWM pump powers and optical fiber lengths. Fig.~\ref{sfig2_2}a is calculated with the joint spectral intensity from Fig.~\ref{sfig2_1}a where the fiber length is set to 200 m. The simulated HOM-dip visibilities are 90.5\%, 95.6\%, and 92.9\% for the BS-FWM pump powers of 4.74 W, 5.20 W, and 5.76 W, respectively. Fig.~\ref{sfig2_2}b is based on the joint spectral intensity from Fig.~\ref{sfig2_1}b, where the fiber length is set to 50 m. The simulated HOM-dip visibilities are 93.7\%, 99.0\%, and 96.2\% for the BS-FWM pump powers of 4.74 W, 5.20 W, and 5.76 W, respectively. As seen from Fig.~\ref{sfig2_2}, the joint-spectral intensity of the 50-m optical fiber shows a more symmetric pattern compared to its 200-m counterpart. This symmetry elevates the visibilities, as a result of the more balanced interference.

\subsection{Frequency-domain NOON-state interference}
The NOON-state interference is experimentally implemented via three steps. First, the NOON state is prepared with the frequency-domain HOM effect via the BS-FWM effect. Then, the relative phase between the NOON state is introduced by the delay line (DL2). Finally, the frequency modes of the NOON state are mixed by another BS-FWM effect, leading to the observation of the NOON-state interference. Starting from the state in equation (\ref{eqS4}), which describes the HOM effect in the frequency domain, the introduced delay $\Delta t_2$ transforms the state as

\begin{equation}
\label{eqS11}
    \begin{aligned}
        \ket{\psi_|out,2|} = \sum_|k| r_|k| & \left( \int d\omega_|s| T(\omega_|s|, \omega_|s0|, \Delta t_|2|) F_|ss,k|(\omega_|s|) \hat{a}^\dagger_|s|(\omega_|s|) + T(\omega_|s|, \omega_|i0|, \Delta t_|2|) F_|si,k|(\omega_|s|) \hat{a}^\dagger_|i|(\omega_|s|) \right) \\
        \times & \left( \int d\omega_|i| T(\omega_|i|, \omega_|s0|, \Delta t_|2|) G_{\rm{is,\Delta t_1,k}}(\omega_|i|) \hat{a}^\dagger_|s|(\omega_|i|) + T(\omega_|i|, \omega_|i0|, \Delta t_|2|)  G_{\rm{ii,\Delta t_1,k}}(\omega_|i|) \hat{a}^\dagger_|i|(\omega_|i|) \right) \ket{0},
    \end{aligned}
\end{equation}

\noindent
where the temporal delay $T(\omega_|x|, \omega_|y0|, \Delta t_|2|) = \exp[i(\omega_|x| + \omega_|y0|) \Delta t_|2|]$. The effect of the other BS-FWM is described as 

\begin{equation}
\label{eqS12}
    \begin{aligned}
    & \ket{\psi_|out,3|} = \sum_|k| r_|k| \left( \int d\omega_1 A'_|k|(\omega_1) \hat{a}^\dagger_|s|(\omega_|1|) + B'_|k|(\omega_|1|) \hat{a}^\dagger_|i|(\omega_|1|) \right) \times
    \left( \int d\omega_|2| C'_|k|(\omega_|2|) \hat{a}^\dagger_|s|(\omega_2) + D'_|k|(\omega_|2|) \hat{a}^\dagger_|i|(\omega_2) \right) \ket{0}, \\
    & A'_|k|(\omega_|1|) = \int d\omega_|s| G'_|ss|(\omega_|s|, \omega_|1|) T(\omega_|s|, \omega_|s0|, \Delta t_|2|) F_|ss,k|(\omega_|s|) + G'_|is|(\omega_|s|, \omega_|1|) T(\omega_|s|, \omega_|i0|, \Delta t_|2|) F_|si,k|(\omega_|s|), \\
    & B'_|k|(\omega_|1|) = \int d\omega_|s| G'_|si|(\omega_|s|, \omega_|1|) T(\omega_|s|, \omega_|s0|, \Delta t_|2|) F_|ss,k|(\omega_|s|) + G'_|ii|(\omega_|s|, \omega_|1|) T(\omega_|s|, \omega_|i0|, \Delta t_|2|) F_|si,k|(\omega_|s|), \\
    & C'_|k|(\omega_|2|) = \int d\omega_|i| G'_|ss|(\omega_|i|, \omega_|2|) T(\omega_|i|, \omega_|s0|, \Delta t_|2|) G_{\rm{is,\Delta t_1,k}}(\omega_|i|) + G'_|is|(\omega_|i|, \omega_|2|) T(\omega_|i|, \omega_|i0|, \Delta t_|2|)  G_{\rm{ii,\Delta t_1,k}}(\omega_|i|), \\
    & D'_|k|(\omega_|2|) = \int d\omega_|i| G'_|si|(\omega_|i|, \omega_|2|) T(\omega_|i|, \omega_|s0|, \Delta t_|2|) G_{\rm{is,\Delta t_1,k}}(\omega_|i|) + G'_|ii|(\omega_|i|, \omega_|2|) T(\omega_|i|, \omega_|i0|, \Delta t_|2|)  G_{\rm{ii,\Delta t_1,k}}(\omega_|i|).
    \end{aligned}
\end{equation}

\noindent
Note that $G'_|xy|(\omega_1, \omega_2)$ is the different Green function from $G_|xy|(\omega_1, \omega_2)$ as the power of the BS-FWM pump for the second stage may be different. By substituting equations (\ref{eqS9}, \ref{eqS10}) with the parameters $A(\omega) = A'(\omega)$, $B(\omega) = B'(\omega)$, $C(\omega) = C'(\omega)$, and $D(\omega) = D'(\omega)$, we theoretically calculate the NOON-state interference in the frequency domain.

\begin{figure}[h]
\centering
\includegraphics[width=6.7in]{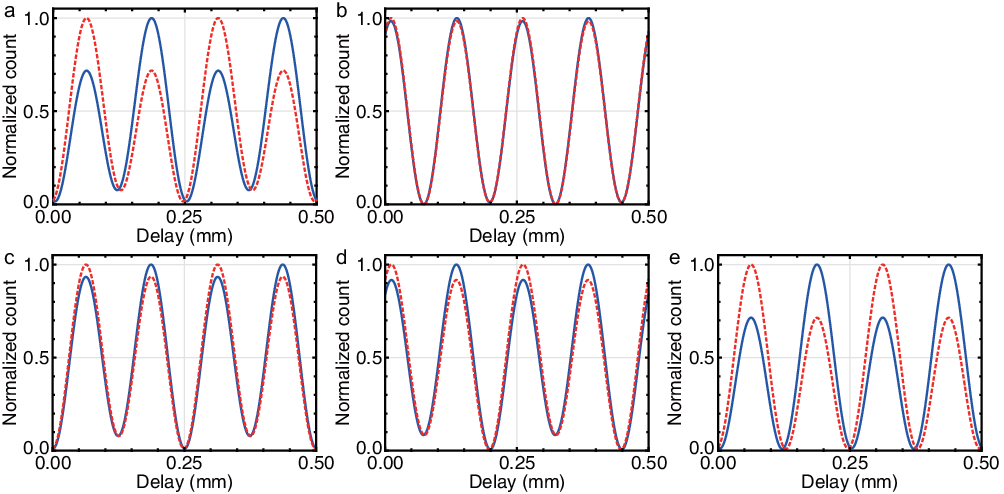}
\caption{\textbf{Simulated frequency-domain NOON-state interference.}
NOON-state interference patterns under various conditions. \textbf{a}, The experimental condition. \textbf{b}, Mitigating the three effects. \textbf{c}, Reducing asymmetry in the two-photon state. \textbf{d}, Compensating walk-off between the two photons. \textbf{e}, Balancing BS-FWM pump powers.
}
\label{sfig2_3}
\end{figure}

Fig.~\ref{sfig2_3} illustrates the frequency-domain NOON-state interference patterns across various simulated conditions. The blue solid curve and red dashed curve denote the normalized counts for the signal and idler modes, respectively. Fig.~\ref{sfig2_3}a presents the interference patterns under our experimental condition. We identify three primary contributors to the observed non-sinusoidal patterns and non-ideal visibilities: 1) asymmetry in the generated two-photon state, 2) walk-off between the two photons in the NZDSF, 3) unbalanced BS-FWM pump powers, 5.76 W for the first stage and 4.74 W for the second. Fig.~\ref{sfig2_3}b represents the interference patterns after rectifying the above three effects. Here, the fitted visibilities of the fringes are 99.1\% for the signal mode and 99.0\% for the idler mode. The three effects are mitigated by the following procedures: 1) Reducing the length of the single-mode fiber from 200 m to 50 m, thereby achieving a symmetrical two-photon state as seen from Fig.~\ref{sfig2_1}, 2) Introducing a 3.23 ps delay to the idler photon after the first-stage BS-FWM process, 3) Equalizing BS-FWM pump powers to 5.20 W for both stages. Fig.~\ref{sfig2_3}(c-e) presents the interference patterns while individually addressing the effects: c) Compensating for the two-photon state asymmetry, d) Canceling the walk-off effect, e) Balancing the BS-FWM pump powers. Therefore, we successfully simulate the frequency-domain NOON-state interference patterns and explain the non-sinusoidal patterns and non-ideal visibilities. The three influential factors for the non-ideal patterns are the asymmetry in the generated two-photon state, the walk-off effect, and the unbalanced BS-FWM pump powers.

\subsection{Frequency-domain single-photon inteference}
The single-photon interference is implemented by heralding the idler photon while the signal photon evolves in the interferometer. As the signal and idler photons propagate in the different path modes, equation (\ref{eqS1}) is modified as 

\begin{equation}
\label{eqS13}
    \ket{\psi_|in,2|} = \sum_|k| r_|k| \int d\omega_|s| F_|k|(\omega_|s|) \hat{a}^\dagger_|s1|(\omega_|s|) \int d\omega_|i| G_|k|(\omega_|i|) \hat{a}^\dagger_|i2|(\omega_|i|) \ket{0},
\end{equation}

\noindent
where $\hat{a}^\dagger_|s1|$ indicates the photon creation operator at the signal frequency mode with the path mode 1, and $\hat{a}^\dagger_|i2|$ is the creation operator at the idler frequency mode with the path mode 2. After the signal photon evolves in the interferometer, the final state is expressed as

\begin{equation}
\label{eqS14}
    \ket{\psi_|out,4|} = \sum_|k| r_|k| \left( \int d\omega_1 A'_|k|(\omega_1) \hat{a}_|s1|^\dagger(\omega_|1|) + B'_|k|(\omega_|1|) \hat{a}^\dagger_|i1|(\omega_|1|) \right) \times \int d\omega_2 G_|k|(\omega_2) \hat{a}^\dagger_|i2|(\omega_2) \ket{0}.
\end{equation}

\noindent
where $A'_|k|(\omega)$ and $B'_|k|(\omega)$ are equivalent with Equation (\ref{eqS12}). The two-photon detection of the signal photon at the path mode 1 and time $t_1$, and the idler photon at the path mode 2 and time $t_2$, is represented as

\begin{equation}
\label{eqS15}
    \hat{E}_|s1|(t_1) \hat{E}_|i2|(t_2) \ket{\psi_|out,4|} = \sum_|k| r_|k| \int d\omega_|1| \exp[-i \omega_|1| t_|1|] A'_|k|(\omega_|1|) \int d\omega_|2| \exp[-i \omega_|2| t_|2|] G_|k|(\omega_2) \ket{0},
\end{equation}

\noindent 
and the probability density of the two-photon detection is calculated as follows,

\begin{equation}
\label{eqS16}
    ||\hat{E}_|s1|(t_1) \hat{E}_|i2|(t_2) \ket{\psi_|out,4|}|^2 = \sum_|k,k'| r_|k| r_|k'| \int d\omega'_|1| d\omega'_|2| d\omega_|1| d\omega_|2| A'^*_|k'|(\omega'_|1|) G^*_|k'|(\omega'_|2|) A'_|k|(\omega_|1|) G_|k|(\omega_|2|) \exp[i(\omega_|1| - \omega'_|1|)t_1 + i(\omega_|2| - \omega'_|2|)t_2].
\end{equation}

\noindent
The time-averaged probability, which is proportional to the measured coincidence count, of the photons at the signal mode at path mode 1 and the idler mode at path mode 2 is calculated as

\begin{equation}
\label{eqS17}
    \int dt_1 dt_2 ||\hat{E}_|s1|(t_1) \hat{E}_|i2|(t_2) \ket{\psi_|out,4|}|^2 = \sum_|k,k'| r_|k| r_|k'| \int d\omega_1 A'^*_|k'|(\omega_1) A'_k(\omega_1) \int d\omega_2 G^*_|k'|(\omega_2) G_|k|(\omega_2) = \sum_|k| r_|k|^2 \int d\omega_1 |A'_|k|(\omega_1)|^2,
\end{equation}

\noindent
where the orthogonality between the Schmidt modes leads to a relation, $\int d\omega G^*_|k'|(\omega) G_|k|(\omega) = \delta_|k,k'|$. With the same procedures, the time-averaged probability of the idler mode at the path mode 1 and the idler mode at the path mode 2 is calculated as follows,

\begin{equation}
\label{eqS18}
    \int dt_1 dt_2 ||\hat{E}_|i1|(t_1) \hat{E}_|i2|(t_2) \ket{\psi_|out,4|}|^2 = \sum_|k,k'| r_|k| r_|k'| \int d\omega_1 B'^*_|k'|(\omega_1) B'_k(\omega_1) \int d\omega_2 G^*_|k'|(\omega_2) G_|k|(\omega_2) = \sum_|k| r_|k|^2 \int d\omega_1 |B'_|k|(\omega_1)|^2.
\end{equation}

\begin{figure}[h]
\centering
\includegraphics[width=6.7in]{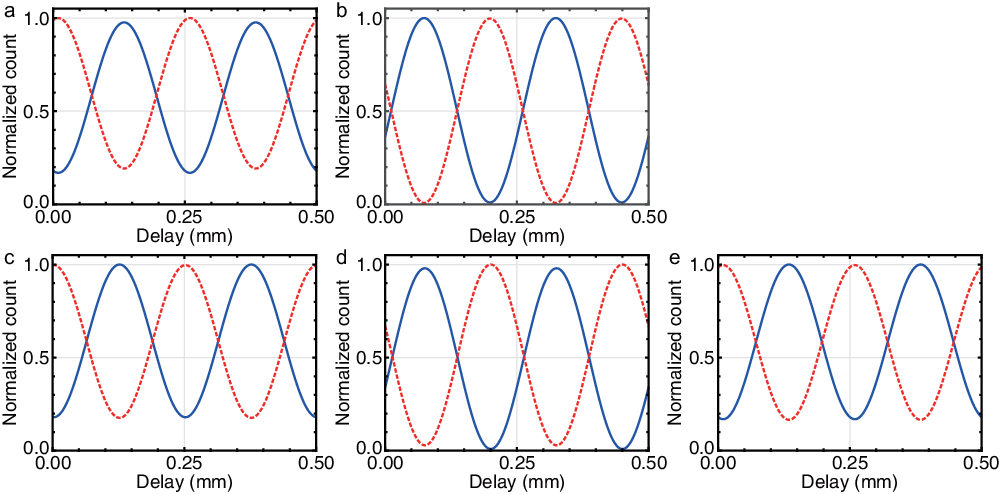}
\caption{\textbf{Simulated frequency-domain single-photon interference.}
Single-photon interference patterns under various conditions. \textbf{a}, The experimental condition. \textbf{b}, Mitigating the three effects. \textbf{c}, Reducing asymmetry in the two-photon state. \textbf{d}, Compensating walk-off between the two photons. \textbf{e}, Balancing BS-FWM pump powers.
}
\label{sfig2_4}
\end{figure}

Fig.~\ref{sfig2_4} depicts the frequency-domain single-photon interference patterns under various simulated conditions. The blue solid curve and red dashed curve denote the normalized counts for the signal and idler modes, respectively. Fig.~\ref{sfig2_4}a presents the interference patterns under our experimental condition. Similar to the previous discussion, three primary factors contribute to the observed non-ideal visibilities: 1) asymmetry in the single-photon spectrum (signal photon), 2) walk-off between the two photons within the NZDSF, 3) Unbalanced BS-FWM pump powers, 5.76 W for the first stage and 4.74 W for the second. Notably, the asymmetry in the two-photon state leads to asymmetry in the single-photon spectrum. Fig.~\ref{sfig2_4}b presents the interference patterns after mitigating the above three effects, where the fitted visibilities of the fringes are 98.1\% for the signal mode and 98.6\% for the idler modes. The three effects are rectified by the following methods: 1) Reducing the length of the single-mode fiber from 200 m to 50 m, which symmetrize the single-photon spectrum. 2) Introducing a 3.23 ps delay to the idler photon after the first-stage BS-FWM process, 3) Balancing BS-FWM pump powers to 5.20 W for both stages.  Fig.~\ref{sfig2_4}(c-e) presents the interference patterns by individually addressing the effects: c) Reducing the asymmetry in the single-photon spectrum, d) Compensating the walk-off effect, e) Balancing the BS-FWM pump powers. Therefore, we simulate the frequency-domain single-photon interference patterns and explain the non-ideal visibilities. The three influential factors for the non-ideal visibilities are the asymmetry in the single-photon spectrum, the walk-off effect, and the unbalanced BS-FWM pump powers.

\clearpage
\section{Frequency-domain HOM interference without accidental subtraction}

\begin{figure}[h]
\centering
\includegraphics[width=3.35in]{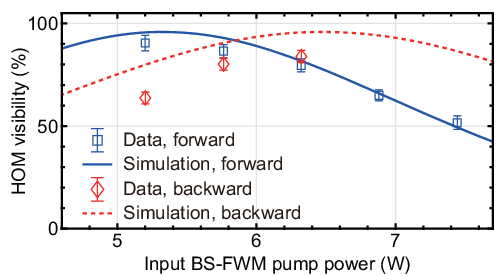}
\caption{\textbf{Frequency-domain HOM interference without accidental subtraction.}
Raw visibility of the HOM interference against the input BS-FWM power. The blue squares and red diamonds represent the raw visibilities of the HOM effect for the forward and backward directions, respectively, and the blue solid and red dashed curves indicate the corresponding simulation results.
}
\label{sfig3}
\end{figure}

In Fig.~\ref{sfig3}, we illustrate the visibilities of the frequency-domain HOM interference as a function of the input BS-FWM power, which determines the splitting ratio of the frequency beam splitter. The blue squares and the red diamonds represent the raw visibilities for the forward and backward directions of the BS-FWM processes, respectively. The error bars are derived from the fitting coefficients and confidence bounds of the measured HOM dips, where the fitting function is denoted as a Gaussian function multiplied by a sinc function. The blue solid and red dashed curves depict the simulated HOM visibilities for the forward and backward directions, respectively. Although the experimental result without the accidental subtraction shows a discrepancy with the simulation result, the overall trends are well consistent.

\clearpage
\section{Fisher information}
While we observe super-resolution, a two-fold enhancement compared to single-photon interference in Figure~5, achieving super-sensitivity remains elusive due to high optical loss in our experimental setup.

In the quantum estimation theory, the lower bound of phase uncertainty, $\Delta\phi_|low|$, is linked to quantum Fisher information, $F_|Q| (\phi)$, by the relation $\Delta\phi_|low| = 1/\sqrt{q F_|Q|(\phi)}$, where ‘$q$’ is the number of repeated measurements~\cite{braunstein1994statistical, fisher1925theory}. For the NOON state, quantum Fisher information is given by $F_|Q| (\phi)= N^2 V^2 \eta_|sys,coincidence| \eta_|g|$, where $V$ is the visibility of the interference fringe, $\eta_|sys,coincidence|$ is the system efficiency, and $\eta_|g|$ is the generation efficiency of the NOON state~\cite{shin2013enhancing}. Under our experimental conditions, the visibility, system efficiency, and generation efficiency are measured at 67\%, 0.0039\%, and 100\%, respectively. This leads to a calculated quantum Fisher information of $7.0 \times 10^{-5}$, resulting in a lower bound of $\Delta \phi_|low| = 120$ for a single-shot measurement ($q = 1$). Since the lower bound exceeds the phase uncertainty of the single-photon interference, $\Delta\phi_|low,single| = 1/(V\sqrt{\eta_|sys,single| \eta_|g|}) = 13$, achieving supersensitivity is currently unattainable. In the single-photon interference experiment, the visibility, system efficiency ($\eta_|sys,single|$), and generation efficiency are 70\%, 1.2\%, and 100\%, respectively.

We expect that the high optical loss can be significantly improved through customizing optical components or implementing the experimental setup on a single chip, such as Silicon or Silicon-nitride platforms. However, we believe that this is beyond the scope of our current paper.

\clearpage
\section{Frequency-domain NOON-state interference without accidental subtraction}

\begin{figure}[h]
\centering
\includegraphics[width=3.35in]{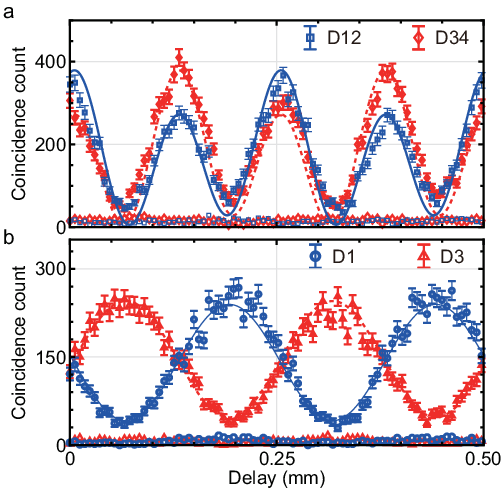}
\caption{\textbf{Frequency-domain NOON-state interference without accidental subtraction.} 
\textbf{a}, Coincidence counts against the delay (DL2) for the NOON-state interference (N = 2). Blue squares and red diamonds with error bars represent coincidence counts between the SNSPDs D1 \& D2 and D3 \& D4, respectively, where markers without error bars indicate the accidental coincidence counts. Solid blue and dashed red curves are simulation results. The error bars are calculated assuming Poissonian statistics of the detection. \textbf{b}, Coincidence counts against the delay (DL2) for the single-photon interference. Blue circle and red triangles with error bars indicate the coincidence counts between the SNSPDs D1 \& D5 and D3 \& D5, respectively, where markers without error bars depict the accidental coincidence counts. 
}
\label{sfig4}
\end{figure}

Fig.~\ref{sfig4} represents the frequency-domain NOON-state interference and single-photon interference without accidental subtraction. Fig.~\ref{sfig4}a depicts the NOON-state interference. Blue squares and red triangles, accompanied by error bars, indicate measured coincidence counts between the SNSPDs D1 \& D2 and D3 \& D4, respectively. In contrast, markers without error bars present accidental coincidence counts. Each data point is measured for 60 seconds. Solid blue and dashed red curves are the simulation results, and the error bars are calculated assuming Poissonian statistics of the detection. Fig.~\ref{sfig4}b represents the single-photon interference. Blue squares and red triangles, with error bars, indicate measured coincidence counts between the SNSPDs D1 \& D5 and D3 \& D5, respectively. Markers without error bars present accidental coincidence counts. Each point is measured for a second. The experimental and simulation results are well consistent. 

\newpage

\bibliographystyle{naturemag}
\bibliography{reference}